\documentclass[journal]{IEEEtran}

\usepackage{subfigure}
\usepackage{epsf}
\usepackage{amsmath,amssymb}
\usepackage{graphicx}
\usepackage{color}
\usepackage{cite}
\usepackage{multirow,tabularx}
\usepackage{ifthen}
\usepackage{times}
\usepackage{stackrel}
\usepackage{amsfonts}
\usepackage{stfloats}
\usepackage[all]{xy}

\input{epsf.sty}

\newcommand{\hide}[1]{\ifthenelse{\boolean{false}}{#1}{}}

\newcommand{\qed}{\nobreak \ifvmode \relax \else
      \ifdim\lastskip<1.5em \hskip-\lastskip
      \hskip1.5em plus0em minus0.5em \fi \nobreak
      \vrule height0.75em width0.5em depth0.25em\fi}

\newcommand{\abs}[1]{\lvert {#1} \rvert}

\newcommand{\bsp}{\begin{slide*}}
\newcommand{\esp}{\end{slide*}}
\newcommand{\bsl}{\begin{slide}}
\newcommand{\esl}{\end{slide}}

\newcommand{\beq}{\begin{equation}}
\newcommand{\eeq}{\end{equation}}
\newcommand{\ben}{\begin{enumerate}}
\newcommand{\een}{\end{enumerate}}
\newcommand{\ber}{\begin{eqnarray}}
\newcommand{\eer}{\end{eqnarray}}

\newcommand{\beqa}{\begin{eqnarray}}
\newcommand{\eeqa}{\end{eqnarray}}

\newtheorem{thm}{Theorem}

\newtheorem{lem}{Lemma}

\newcommand{\lb}{\left (}
\newcommand{\rb}{\right )}

\newcommand{\norm}[1]{\Vert#1\Vert}

\newcommand{\mean}{\mathbb{E}}

\newcommand{\yAtrain}{{y}_{A,\tau}}
\newcommand{\yBtrainvec}{\mathbf{y}_{B,\tau}}
\newcommand{\yBtrainvectwo}{\mathbf{y}_{B,\tau_2}}

\newcommand{\yBtrainvectildetwo}{\tilde{\mathbf{y}}_{B,\tau_2}}
\newcommand{\yBdatavec}{\mathbf{y}_{B,d}}
\newcommand{\xAtrain}{{x}_{A,\tau}}
\newcommand{\xAtraintwo}{{x}_{A,\tau_2}}
\newcommand{\xBtrainvec}{\mathbf{x}_{B,\tau}}
\newcommand{\xAdata}{{x}_{A,d}}
\newcommand{\channel}{\mathbf{h}}

\newcommand{\barp}{\bar{P}}
\newcommand{\LAtone}{L_{A,\tau_1}}
\newcommand{\LAttwo}{L_{A,\tau_2}}
\newcommand{\LB}{L_{B,\tau}}

\newcommand{\noisevecBt}{\mathbf{w}_{B,\tau}}
\newcommand{\noisevecBttwo}{\mathbf{w}_{B,\tau_2}}
\newcommand{\noisescalarAt}{{w}_{A,\tau}}
\newcommand{\noisescalarAtbar}{\bar{w}_{A,\tau}}

\newcommand{\noisevecBd}{\mathbf{w}_{B,d}}

\newcommand{\kappone}{\kappa_{\bar P}}

\newcommand{\lcoh}{L_c}
\newcommand{\vhat}{\hat{\mathbf{v}}}
\newcommand{\singvec}{{\mathbf{v}}}
\newcommand{\singval}{{\sigma}}
\newcommand{\standgauss}{{\mathcal{CN}(0,1)}}
\newcommand{\complex}{\mathcal{C}}
\newcommand{\ptr}{\mathcal{P}(\hat{\sigma})}
\newcommand{\pcontrol}{\frac{\exp\left(\frac{R_{\barp} \lcoh}{\lcoh-\LB}\right)-1}{\singval^2}}
\newcommand{\ptrtrain}{\mathbf{p}_c}
\newcommand{\ptrtrainhat}{\hat{\mathbf{p}}_c}
\newcommand{\ptrtraintilde}{\tilde{\mathbf{p}}_c}
\newcommand{\setxysize}{\epsfysize=2.2in \epsfxsize=2.8in}

\IEEEoverridecommandlockouts

\begin{document}

\title{On the DMT of TDD-SIMO Systems with Channel-Dependent Reverse Channel Training}
 \author{B. N. Bharath and Chandra R. Murthy {\it Senior Member, IEEE}

\thanks{The authors are with the Dept.\ of Electrical Communication
     Eng.\ at IISc, Bangalore,
     India. (e-mails: \{bharath, cmurthy\}@ece.iisc.ernet.in)}

\thanks{The authors gratefully acknowledge the financial support of the DST, India and of the EPSRC, UK under the auspices of the India-UK Advanced Technology Center (IU-ATC), as well as that of the Defense Research Development Organization, India.}
\thanks{This work has appeared in part in \cite{bharath2010improvement}.}

}
\maketitle

\newcommand{\ceil}[1]{\lceil{#1}\rceil}
\newcommand{\subbandsel}[1]{{\cal I}_{#1}}
\newcommand{\Xeff}{X_{\text{eff}}}

\begin{abstract}
This paper investigates the Diversity-Multiplexing gain Trade-off (DMT) of a training based reciprocal Single Input Multiple Output (SIMO)  system, with (i) perfect Channel State Information (CSI) at the Receiver (CSIR) and noisy CSI at the Transmitter (CSIT), and (ii) noisy CSIR and noisy CSIT. In both the cases, the CSIT is acquired through Reverse Channel Training (RCT), i.e., by sending a training sequence from the receiver to the transmitter. A channel-dependent fixed-power training scheme is proposed for acquiring CSIT, along with a forward-link data transmit power control scheme. With perfect CSIR, the proposed scheme is shown to achieve a diversity order that is quadratically increasing with the number of receive antennas. This is in contrast with conventional orthogonal RCT schemes, where the diversity order is known to saturate as the number of receive antennas is increased, for a given channel coherence time. Moreover, the proposed scheme can achieve a larger DMT compared to the orthogonal training scheme. With noisy CSIR and noisy CSIT, a three-way training scheme is proposed and its DMT performance is analyzed. It is shown that nearly the same diversity order is achievable as in the perfect CSIR case. The time-overhead in the training schemes is explicitly accounted for in this work, and the results show that the proposed channel-dependent RCT and data power control schemes offer a significant improvement in terms of the DMT, compared to channel-agnostic orthogonal RCT schemes. The outage performance of the proposed scheme is illustrated through Monte-Carlo simulations.
\end{abstract}
\begin{keywords}
Diversity-multiplexing gain tradeoff, MMSE channel estimation, training sequence.
\end{keywords}

\section{Introduction} \label{sec:intro}
Reliability and system throughput are two fundamental parameters of interest in any wireless communication system, and the inherent tradeoff between the two at high SNR was elegantly captured by the Diversity Multiplexing gain Tradeoff (DMT) proposed in the seminal work of Zheng and Tse \cite{zheng2003diversity}.
It is known that a significant improvement in the outage performance can be obtained if the Channel State Information (CSI) at the receiver (CSIR) and the transmitter (CSIT) are perfect \cite{biglieri2001limiting}, \cite{caire1999optimum}, while \cite{zheng2003diversity} considered perfect CSIR and no CSIT.

In a Time Division Duplex (TDD) system, CSI could be estimated at the transmitter and receiver by sending a known training sequence in the forward and reverse-link directions, respectively. This has two consequences. First, the \emph{estimation error} results in incorrect  data rate or power adaptation at the transmitter, in turn leading to higher outage rate. Second, training incurs a \emph{time overhead}, which could be non-trivial when the training occupies a significant fraction of the channel coherence time, as it affects the pre-log term in the achievable data rate~\cite{ref3}.
This paper therefore focuses on the important problem of analytically comparing the DMT performance of different channel estimation techniques and identifying training signals and data power control schemes that result in a good performance in terms of the achievable DMT. We start with a brief survey of related literature. 

The impact of imperfect CSIT on the DMT of a multiple antenna system has been a popular area of research, and it is known that
even with imperfect CSIR and CSIT, a significant improvement in DMT can be obtained, compared to the no-CSIT case (see, for example, \cite{ElGamalCaireDamenTITAug2006, KhoshnevisSabharwalAllerton2004, RaghavaSharmaAllerton2005}). The effect of imperfect CSIR on the DMT of a MIMO system was first studied in \cite{zheng2002diversity}.
The DMT analysis of a multiple antenna system with perfect CSIR and when the CSIT is modeled as the CSI plus Gaussian noise whose variance decreases with training SNR was investigated in \cite{lim2006fundamental, lim2008fundamental, KimCaireTITApr2009}. In a TDD setup, the achievable DMT improvement using power control based on noisy CSIT was shown in \cite{StegerSabharwalTWCDec2008, KimCaireTITApr2009, ZhangGongLetaiefTCOMApr2010}.
Other works that study the DMT performance with quantized feedback of CSI and/or target data rate control based on noisy CSIT include \cite{KhoshnevisSabharwalAllerton2004, KhoshnevisSabharwalICC2004, KimSkoglundTITAug2007, ElGamalCaireDamenTITAug2006, lim2008fundamental, AggarwalSabharwalTITJune2011, ZhangGongLetaiefTWCJan11}. In \cite{AggarwalSabharwalTITJune2011, ZhangGongLetaiefTWCJuly11}, the DMT of two-way and multi-round training schemes in a TDD system was derived. In these studies, the channel feedback signal on the reverse link is chosen to satisfy an average power constraint, rather than an instantaneous power constraint. 

Most of the aforementioned studies of the DMT with imperfect CSI typically ignore the training duration overhead. Hence, they are primarily applicable to slowly varying channels, where the time overhead in training occupies an insignificant fraction of the channel coherence time. An exception is  \cite{StegerSabharwalTWCDec2008}, where, taking the training overhead into account, the authors concluded that for nonzero multiplexing gain $g_m$, the diversity order saturates as $r$ increases, where $r$ is the number of receive antennas.
Hence, for fast varying channels, the authors suggest turning off receive antennas in order to achieve higher multiplexing gains.
It is important to account for the training duration overhead in deriving the achievable DMT, because, as the SNR goes to infinity, although the estimation error goes to zero, the training duration overhead remains fixed and has a direct impact on the DMT.
Also, by modeling the CSIT as the sum of the true CSI and an additive error, most of the past studies implicitly assume that a channel-agnostic orthogonal training signal is employed for channel estimation. When the training signal is channel-dependent, the imperfect CSI can no longer be modeled as the sum of the true CSI and an additive noise. Due to this, the existing results cannot be directly extended to analyze the DMT performance of channel-dependent training schemes.

When the channel is reciprocal and block-fading, e.g., in a TDD system, the receiver could exploit its channel knowledge (acquired through an initial forward-link training phase) in designing its reverse-training sequence, not only to reduce the channel estimation error at the transmitter, but also to reduce the required training duration overhead.
Hence, the goals of this paper are two-fold: (a) to analyze the DMT performance of a \emph{channel dependent} training scheme for acquiring CSIT and an associated power control mechanism for data transmission;  and (b) to contrast the DMT performance of the proposed training and power control schemes with that achieved by conventional channel agnostic training schemes. Our study focuses on point-to-point Single Input Multiple Output (SIMO) systems. This is of  practical importance, since it applies, for example, to the uplink of wireless networks where the base station has multiple antennas, the mobile users have a single antenna, and orthogonal access is used (e.g., OFDM/TDMA) as in WLANs and 4G/LTE systems.  The channel dependent training sequence employed here was first proposed by us in \cite{BharathICASSP2009} and \cite{bharath2010improvement} in a MIMO and SIMO context, respectively, and was independently explored in \cite{zhou2010two}, although not in a DMT context. 

In this paper, for analytical simplicity and clarity of presentation, we start by assuming that perfect CSI is available at the receiver, as in \cite{lim2006fundamental, lim2008fundamental, KimCaireTITApr2009}.
 We propose a fixed-power RCT sequence, using which, the CSI can be estimated at the transmitter using a minimum duration of only one symbol, i.e., with a \emph{factor of $r$ reduction} in training duration compared to orthogonal RCT. 
For data transmission, we propose a modified truncated channel inversion-type power control scheme based on the noisy CSIT.
For this system, we show that a diversity of $d(g_m) = r\left(s+1-\frac{g_m \lcoh}{\lcoh-\LB}\right)$ is achievable. Here, $g_m$ is the multiplexing gain, $\lcoh$ is the coherence time, $\LB \ge 1$ is the reverse training duration, and $1 \le s < r$ is a parameter in the data power control scheme.~(See Section~\ref{sec:syspftcsir}.)

Next, we consider the more practical case where noisy CSIR is acquired via a forward link training sequence, and propose a \emph{three-way} training scheme followed by data transmission. We show that a DMT of  $d(g_m) = r(s+1-\frac{g_m \lcoh}{\lcoh-\beta})$ is achievable, where $\beta \geq 3$ is the total training overhead from all three training phases, which is again an improvement over conventional orthogonal training schemes. For example, a nonzero diversity order can be achieved with $\frac{\lcoh - (r+2)}{\lcoh} \le g_m < \frac{\lcoh - 3}{\lcoh}$, which is not possible with orthogonal training schemes without switching off receive antennas and incurring an associated reduction in diversity order.~(See Section~\ref{sec:twoway}.)

Note that although the perfect CSIR case is a special case of the three-way training scheme with infinite forward-link training power, we briefly present the perfect CSIR case also, as it provides insights into the impact of the reverse-training and data power control mechanisms on the DMT. Moreover, it is useful as an upper bound on the performance with imperfect CSIR.
Also, we assume that power control is employed only at the transmitter and focus on fixed-power RCT in the sequel. Using power controlled RCT significantly changes the problem; we analyze this case in our follow up work \cite{Bharath2011UnderPrep}.

An important implication of our work is that it shows that by exploiting the receiver's knowledge of the CSI in designing the \emph{reverse channel training (RCT)} sequence and using our proposed data power control scheme, one can achieve a higher diversity order than conventional RCT for all values of $g_m$.  Somewhat surprisingly, we also demonstrate that although the DMT analysis corresponds to taking the SNR to infinity, it can nonetheless be used to discriminate between different training schemes both in terms of the estimation error as well as the training overhead. At finite SNR, this translates to an improvement in the outage probability performance and the achievable data rate, as will be illustrated through Monte-Carlo simulations in Section~\ref{sec:sims}.

We use the following notation. Bold face letters are used for vectors and normal font letters are used for scalars. We use $\mathbb{E}(\cdot)$ to denote the expected value of $(\cdot)$. 
We use $\norm{\channel}_2$ to represent the $\ell_2$ norm of $\channel$. The transpose conjugate, absolute value, and real part are denoted by $(\cdot)^H$, $|\cdot|$ and $\Re\{\cdot\}$, respectively.
We write $f(\barp) \doteq \frac{1}{\barp^k}$ to mean $-\lim_{\barp \rightarrow \infty}{\frac{\log f(\barp)}{\log{\barp}}} = k.$ Similarly, we define $f(\barp) \preceq \frac{1}{\barp^k}$ to mean $-\lim_{\barp \rightarrow \infty}{\frac{\log f(\barp)}{\log{\barp}}} \geq k$.

\section{System Model}\label{sec:sysmod}
The system model consists of two communicating nodes, \emph{node A} with a single antenna and \emph{node B} with $r$ antennas, with \emph{node A} attempting to send data to \emph{node B} over a wireless channel. The forward channel from \emph{node A} to \emph{node B}, denoted by $\channel \in \complex^{r \times 1}$, is modeled as a Rayleigh flat fading channel whose entries are \emph{i.i.d.} Circularly Symmetric Complex Gaussian (CSCG) random variables with zero mean and unit variance, i.e., $\standgauss$. The channel is assumed to be block-fading, i.e., it remains constant for a duration of the coherence time $\lcoh$, and evolve in an \emph{i.i.d.} fashion across coherence times. We assume a TDD system with perfect reciprocity, and hence, taking the complex conjugate of the received signal at \emph{node A},  the reverse link channel is $\channel^H$. We let $\channel = \singval \singvec$, where $\sigma = \norm{\channel}_2$ is the singular value and $\singvec \triangleq \frac{\channel}{\norm{\channel}_2}$ is the singular vector of $\channel$. Since our goal is to study the achievable DMT performance with channel training, we first explain the two-way training protocol used for acquiring CSI at \emph{node B} and \emph{node A}. Later, in Sec. \ref{sec:twoway}, an additional phase of forward link training is introduced, which is not presented here for simplicity.

\setcounter{subsubsection}{0}
\subsubsection{\textbf{Phase I (Forward-link training)}} \label{sec:phase1train}
Here, the training sequence $\xAtrain = \sqrt{\barp \LAtone}$ is transmitted from \emph{node A} to \emph{node B}, where $\LAtone$ denotes the training duration and ${\barp}$ is the training power\footnote{Strictly speaking, $\xAtrain = \sqrt{\barp}$ is transmitted repeatedly $\LAtone$ times. Mathematically, this is equivalent to using $\xAtrain = \sqrt{\barp \LAtone}$ for a duration of one unit.}. Throughout this paper, we use $\barp$ as the average power constraint during both training and data transmission.
The corresponding received training signal is given by,
\beq
\yBtrainvec = \channel \sqrt{\barp \LAtone} + \noisevecBt.
\eeq
The entries of $\noisevecBt \in \complex^{r \times 1}$ are assumed to be distributed as \emph{i.i.d.} $\standgauss$. From the received training signal $\yBtrainvec$, \emph{node B} computes an MMSE estimate of $\channel$, denoted $\hat{\channel}$. The error in the estimate, denoted $\tilde{\channel} \triangleq \channel - \hat \channel$, has \emph{i.i.d.} $\mathcal{CN}\lb 0, 1/(1 + \barp \LAtone) \rb$ distributed entries.

In a TDD-SIMO system, \emph{node A} only requires knowledge of $\singval$ to perform power control, which in turn improves the diversity order compared to the no-CSIT case. Therefore, in phase~II, we estimate only $\singval$ at \emph{node A}, using a channel dependent training sequence.

\subsubsection{\textbf{Phase II (Reverse-link training)}} \label{sec:phase2train}
Since \emph{node B} has an estimate (say, $\hat{\singvec} \triangleq \frac{\hat{\channel}}{\norm{\hat{\channel}}_2}$) of the channel, in this phase, it exploits its CSI to transmit the following training sequence \cite{bharath2010improvement, BharathICASSP2009}:
\beq \label{eq:tw2}
\xBtrainvec = \sqrt{\barp \LB} \vhat,
\eeq
where $\LB$ is the reverse training duration. Using the corresponding received signal,  $\yAtrain\triangleq \channel^H \xBtrainvec + \noisescalarAt,$ where  the AWGN $\noisescalarAt \in \complex$ is distributed as $\standgauss$, \emph{node A} computes an estimate of the singular value as follows:
\beq \label{eq:tw92}
\hat \singval \triangleq \frac{\Re\{\yAtrain\}}{\sqrt{\barp \LB}} = \singval \Re\{\singvec^H \hat{\singvec}\} +\noisescalarAtbar,
\eeq
where $\noisescalarAtbar \triangleq \frac{\Re\{\noisescalarAt\}}{\sqrt{\barp \LB}}$. Note that the estimate $\hat{\singval}$ could be negative; this is taken care of by the power control proposed in Sec.~\ref{sec:syspftcsir}, which uses $\hat{\singval}$ only when it is greater than a positive threshold. Since a low or negative $\hat{\singval}$ is likely to be inaccurate, the thresholding technique helps to avoid the poor DMT performance due to such estimates.
The RCT scheme employed above is different from existing channel agnostic methods in that the minimum training length in the proposed scheme is only $1$ symbol. This represents a factor of $r$ reduction compared to orthogonal RCT schemes, where the minimum training length increases linearly with $r$, and this difference in overhead could be significant when $\lcoh$ is small. Also, if $\vhat$ is error-free, it is the optimal beamforming vector for estimating $\sigma$ at \emph{node A}. 

\subsubsection{Multiplexing Gain and Diversity Order}
We recall the definitions of the multiplexing gain, $g_m$, and the diversity order $d$ from \cite{zheng2003diversity}:
\beq \label{eq:MULTG}
g_m \triangleq \lim_{\bar{P} \rightarrow \infty}{\frac{R_{\barp}}{\log{\bar{P}}}},~~~~
d \triangleq - \lim_{\bar{P} \rightarrow \infty}{\frac{\log{P_{out}}}{\log{\bar{P}}}},
\eeq
where $R_{\barp}$ is the target data rate when the average data power constraint is $\barp$, and $P_{out}$ is the corresponding outage probability, i.e., the probability that $R_{\barp}$ exceeds the channel capacity. In this work, the target data rate $R_{\barp} = g_m \log \barp$ is fixed and is independent of the CSIT; the extension of our proposed methods to joint rate and power adaptation is relegated to future work. The rate of data transmission $R_{\barp}$ is increased with $\barp$ by increasing the cardinality of the signal set, keeping the symbol duration fixed.  We ignore the effect of spectral leakage, and assume that the signal bandwidth remains fixed as $\barp$ goes to infinity. Also, we use outage probability as a proxy for the probability of error at high SNR with finite-length codes; this is because the probability of error can be made to decrease as fast as the outage probability using finite-length approximately universal codes \cite{TavildarViswanathTITJuly2006, EliaEtAlTITSep2006}.

In the next section, we assume perfect CSI at \emph{node B} and derive the achievable DMT performance of our proposed training and data transmission schemes.

\section{DMT Analysis with Perfect CSIR} \label{sec:syspftcsir}
When the CSIR is perfect, we have $\hat \singvec = \singvec$, and in this case, it is easy to see that \eqref{eq:tw2} is optimal for estimating $\singval$ given a power constraint $\barp$ on the training signal.
This is because, in general, the training signal can be expressed as the linear combination $\xBtrainvec = \delta \singvec + \beta \singvec_{\perp}$, where $\singvec_{\perp}$ is orthogonal to $\singvec$ and $\delta$ and $\beta$ are some constants. Then, the received training signal at \emph{node A} is $\yAtrain = \delta \singval + \noisescalarAt$, i.e., the power in $\singvec_{\perp}$ does not help in estimating $\singval$.  From \eqref{eq:tw92}, an \emph{unbiased} estimator of the singular value at \emph{node A} is given by
\beq \label{eq:SINGEST}
\hat{\singval} = \singval + \noisescalarAtbar.
\eeq
Note that since the channel is assumed to be Rayleigh fading, $\singval^2$ is chi-square distributed with $2r$ degrees of freedom.
Also, we employ this estimator primarily because we are interested in deriving the achievable DMT performance, and for this purpose, this simple unbiased estimator is sufficient.

\subsection{Power-Controlled Data Transmission from \emph{Node~A} to \emph{Node~B}} \label{sec:datatxperfectcsir}
Given the CSIT $\hat \singval$ in \eqref{eq:SINGEST}, \emph{node A} uses a power $\ptr$ in the forward link \emph{data transmission} phase, to avoid outages while satisfying the average data power constraint $\bar{P}$.
The corresponding \emph{data signal} received at \emph{node B} is given by,
\beq \label{eq:DAT}
\yBdatavec = \sqrt{\ptr} \channel \xAdata + \noisevecBd,
\eeq
where $\xAdata \sim \standgauss$, and with appropriate power normalization, the entries of the AWGN $\noisevecBd \in \complex^{r \times 1}$ are assumed to be \emph{i.i.d.} $\standgauss$. Also, $\ptr$ is chosen independent of $\xAdata$ such that $\mean\{\ptr\} = \barp$, where the expectation is with respect to $\hat \singval$ given in \eqref{eq:SINGEST}, taken across all coherence blocks. Since $\mean\{\abs{\xAdata}^2\} = 1$ within a block, this ensures that the average data power constraint at \emph{node A} is satisfied.

\begin{figure}[ht]
  \begin{center}
  \setxysize
    {\includegraphics[width=3.0in]{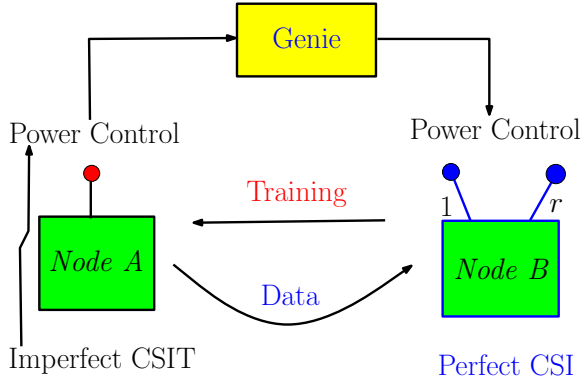}}
  \caption{System model for reverse channel training with perfect CSIR used in Section \ref{sec:syspftcsir}.} \label{fig:m10}
  \end{center}
  \end{figure}

We now present the data power control function $\ptr$ considered in this paper.
Our proposed power control function is motivated as follows. The capacity of a fading channel with \emph{mismatched} CSIT and CSIR is not known in closed form \cite{ref11}. Since the outage probability computation requires a closed form expression for the capacity, we consider a genie-aided receiver as in~\cite{StegerKhoshnevisSabharwalAazhangISIT2007}, where \emph{node B} is assumed to know $\ptr$. This is schematically illustrated in Fig.~\ref{fig:m10}. Then, the achievable data rate
conditioned on the knowledge of $\sqrt{\ptr} \channel$
is given by \cite{ref11}
\beq \label{eq:cap}
C \triangleq \frac{\lcoh-\LB}{\lcoh}\log\left(1+\sigma^2 \ptr \right).
\eeq
An outage occurs when $R_{\barp}$, the target data rate, exceeds $C$. Its probability is upper bounded by
\beq \label{eq:OUT}
P_{out} \triangleq \Pr\left(\frac{\lcoh-\LB}{\lcoh}\log(1+\sigma^2 \ptr) < R_{\barp}\right).
\eeq
Note that the exact outage probability is obtained by minimizing the right hand side above over all $\ptr$ satisfying $\mean\{ \ptr \} = \barp$. Hence, using our proposed data power control scheme leads to an upper bound on the outage probability,
which is sufficient for obtaining the achievable DMT performance.
If the CSIT is perfect (i.e., $\hat{\singval}^2 = \singval^2$),  it is shown in \cite{biglieri2001limiting} that the power control that minimizes the outage probability is given by
\beq \label{eq:optpalloc}
\Phi(\singval^2) \triangleq \pcontrol.
\eeq
Note that since $R_{\barp} = g_m \log \barp$ and $\mean\left\{\tfrac{1}{\singval^2}\right\} = \tfrac{1}{r-1}$, $\Phi(\singval^2)$ satisfies $\mean\{\Phi(\singval^2)\} \le \barp$ for large enough $\barp$, provided $g_m \le (\lcoh - \LB)/\lcoh$. With inaccurate CSIT, due to the estimation error in $\hat{\singval}$, the natural extension of using a transmission power of $\Phi(\hat \singval^2)$ could result in  allocating insufficient power or more power than required, which could lead to suboptimal performance. Also, inverting the channel for all values of $\hat{\singval}$ results in an infinite average power since the Gaussian noise can make the estimate $\hat{\singval}$ arbitrarily small with a non-zero probability. One solution is to use a transmit power of $\Phi(\hat{\singval}^2)$ when $\hat \singval > \theta_0$ and a zero power otherwise, where $\theta_0$ is chosen such that $\mean[\Phi(\hat \singval^2) 1_{\hat \singval > \theta_0}] = \barp$.
The drawback of this method is that it results in an outage probability of $1$ when $\hat{\singval} \le \theta_0$, leading to a zero diversity order. To overcome this problem, we choose the threshold  $\theta_0$ such that $\theta_0 \rightarrow 0$ as $\barp \rightarrow \infty$. Moreover, when $\hat{\singval} \le \theta_0$, we do not necessarily want to use zero power, since the small value of $\hat{\singval}$ could be due to the estimation error. This motivates the following modified power control:
\beq \label{eq:DATRULE}
\ptr \triangleq \left\{
      \begin{array}{ll}
      \bar{P}^l          &  \hat{\singval} \le \theta_{\barp},\\
      \kappone \times \Phi(\hat{\sigma}^{2s}) & \hat{\singval} > \theta_{\barp},
      \end{array}
  \right.
\eeq
where $s \ge 1$ is a parameter, and we use $\theta_{\barp} \triangleq \frac{1}{\barp^n}$, $n>0$, for mathematical tractability.
The parameters $n$, $\kappone$ and $l>0$ are chosen such that $\mean{[\ptr]} = \barp$. Although similar power control schemes have been employed in the literature with perfect CSIT \cite{biglieri2001limiting} or orthogonal RCT \cite{KimCaireTITApr2009, StegerSabharwalTWCDec2008, ZhangGongLetaiefTWCJuly11}, the form in \eqref{eq:DATRULE} is new.
Specifically, the power control scheme in \cite{biglieri2001limiting,KimCaireTITApr2009, StegerSabharwalTWCDec2008} can be obtained from \eqref{eq:DATRULE} by setting $s=1$, $\theta_{\barp} = 0$ and $l = -\infty$; while that in   \cite{ZhangGongLetaiefTWCJuly11} can be obtained by setting  $s= r$, $\theta_{\barp} = 0$ and $l  = -\infty$.

\subsubsection*{Power constraint} 
The description of the power control would be complete if the parameters $n$, $\kappone$ and $l$ can be chosen such that $\mean{[\ptr]} = \barp$, which is the essence of the following Lemma.
\begin{lem} \label{lem:ptrcontrol}
Let $\theta_{\barp} \triangleq \frac{1}{\sqrt{\barp}}$. For $1 \le s < r$, there exists a $\kappone \doteq \frac{1}{\barp^{\frac{g_m}{\alpha} - 1}}$, where  $\alpha \triangleq \frac{\lcoh - \LB}{\lcoh}$, such that $\mean{[\ptr]} = \barp$, if $0\leq l \leq r+1$.
\end{lem}
\emph{Proof}: See Appendix \ref{app:ptrcontrol}. $\blacksquare$

Due to Lemma \ref{lem:ptrcontrol}, in the rest of this paper, we consider $\theta_{\barp} = 1/\sqrt{\barp}$. Also, in Sec.~\ref{sec:twoway}, we show that a minor modification of the above data power control scheme can be employed even with imperfect CSIR.
The next subsection presents the achievable DMT of the proposed training and power control schemes.

\subsection{Achievable DMT Analysis} \label{sec:OUT}
\begin{thm} \label{thm:thmdmt1}
Given $r$ receive antennas and $\LB$ training symbols being used per coherence
interval $\lcoh$ to estimate the CSIT in a SIMO system with perfect CSIR and a genie-aided receiver, an achievable diversity order as a function of multiplexing gain $g_m$ is given by
\beq \label{eq:DMT}
d(g_m) = r \left(\min\{l,s+1\}-\frac{g_m}{\alpha}\right),
\eeq
where $0 \leq l \leq r+1$, $1 \le s < r$, $0 \leq  g_m < \alpha$, and $\alpha \triangleq \frac{\lcoh-\LB}{\lcoh}$ represents the fractional data transmit duration. 
\end{thm}
\emph{Proof:} See Appendix \ref{app:thmdmt1}. $\blacksquare$

\textbf{Remark}:  From a DMT perspective, it is clear from Theorem \ref{thm:thmdmt1} that $s \rightarrow r, l = r+1$ is superior to $s = 1, l=2$. On the other hand, when $\hat \singval < 1$, $\Phi(\hat{\singval}^{2r})$ could be much greater than $\Phi(\hat{\singval}^{2})$.
Thus, in practical systems with a  peak power per transmitted codeword constraint, $s=1, l=2$ could be preferable over $s \rightarrow r, l = r+1$. In the sequel, for convenience, we associate $l=2$ with $s=1$ and $l=r+1$ with $s \rightarrow r$, and drop the explicit dependence of the diversity order on $l$. Further remarks and discussions on the result obtained here are deferred to Sec.~\ref{sec:discussion}.

\section{Three Way Training} \label{sec:twoway}
\begin{table*}[b]
  \centering
  \caption{Three Way Training in a TDD-SIMO System}
   \label{twoway}
  \begin{tabular}{|c|c|c|}
  \hline
 Phase & Description & Input-Output Equation  \\ \hline
  \emph{I} &Fixed power training (\emph{Node A} $\rightarrow$ \emph{Node B}) &$\yBtrainvec = \channel \xAtrain + \noisevecBt$ \\ \hline
  \emph{II} &Fixed power training  (\emph{Node B} $\rightarrow$ \emph{Node A})  &$\yAtrain = \channel^H \xBtrainvec + \noisescalarAt$\\ \hline
  \emph{III} &Power controlled training  (\emph{Node A} $\rightarrow$ \emph{Node B}) &$\yBtrainvectwo =  \hspace{-2mm}\sqrt{\barp \LAttwo \ptr}  \channel + \noisevecBttwo$\\ \hline
  \emph{IV} &Power controlled data  (\emph{Node A} $\rightarrow$ \emph{Node B}) &$\yBdatavec = \channel \xAdata + \noisevecBd$\\ \hline
\end{tabular}
\end{table*}
In this section, we consider the more practical scenario where training is performed in both directions. 
We show that with fixed power training, one can achieve nearly the same DMT
as derived in Sec. \ref{sec:syspftcsir} for the perfect CSIR case.  Unlike in the previous section, the analysis presented here is exact, in the sense that it does not require the assumption of a genie aided receiver, and hence, the DMT derived here is indeed achievable in practice.
The transmission protocol now consists of four phases, as shown in Table \ref{twoway}. The CSIR and CSIT are obtained by transmitting a fixed power training sequence in both directions, as explained in Sec. \ref{sec:sysmod}. However, even a small mismatch in the CSI knowledge at \emph{node A} and \emph{node B} can potentially lead to a large mismatch in their estimate of the data transmit power~\cite{StegerSabharwalTWCDec2008}. Thus, it is essential to train \emph{node B} about \emph{node A}'s knowledge of $\ptr$. This leads to a third phase of training, which is an additional power-controlled forward-link training phase. First, in the following subsection, we explain the power control scheme that is employed here.

\subsection{Power Control Scheme} \label{sec:TWPC}
The power control scheme we propose to employ in this section is as given by \eqref{eq:DATRULE}, due to the following. Let $\hat{\channel}$ denote the MMSE estimate of the channel at \emph{node B}, and consider $\hat \singval$ in \eqref{eq:tw92}. We have
\beqa
\hat \singval \triangleq \frac{\Re\{\yAtrain\}}{\sqrt{\barp \LB}}
\hspace{-2mm}&=& \hspace{-2mm}\Re\{\hat{\channel}^H \hat{\singvec}\}+ {\Re\{\tilde{\channel}^H \hat{\singvec}\} + \frac{\Re\{\noisescalarAt\}}{\sqrt{\barp \LB}}} \nonumber \\
\hspace{-2mm}&=& \hspace{-2mm}\norm{\hat{\channel}}_2 + \tilde{w}_{eff},\label{eq:tw92replicate}
\eeqa
where $\tilde{w}_{eff} \triangleq {\Re\{\tilde{\channel}^H \hat{\singvec}\} + \frac{\Re\{\noisescalarAt\}}{\sqrt{\barp \LB}}}$. Note that $\hat \channel$ and $\tilde \channel$ are independent Gaussian random variables\footnote{$\hat \channel \rightarrow \channel$ as $\barp \rightarrow \infty$. Moreover, $\norm{\hat{\channel}}_2$ is a chi distributed random variable.}. Since $\hat \singvec$ is uniformly distributed on the unit sphere and is independent of $\tilde{\channel}$, $\Re\{\tilde{\channel}^H \hat{\singvec}\}$ is  Gaussian distributed.
This implies that the effective noise, $\tilde{w}_{eff}$, is Gaussian distributed with $\mean{\abs{\tilde{w}_{eff}}^2} \doteq \frac{1}{\barp}$ and independent of $\hat{\channel}$.
Therefore, the estimate of the singular value at \emph{node A} is statistically similar to the estimate given by \eqref{eq:SINGEST} in the perfect CSIR case. Thus, we use a similar power control, $\ptr$ in
\eqref{eq:DATRULE}, where $\hat{\singval}$ is given by~\eqref{eq:tw92replicate}. Also, with a slight abuse of notation, $\alpha \triangleq \frac{\lcoh - \LB - L_{A,\tau_1} - L_{A,\tau_2}}{\lcoh}$, where $L_{A, \tau_2}$ is the training duration in the third phase of training (phase III), which is in the forward-link direction.

In this section, without loss of generality, we move the power scaling $\sqrt{\barp}$ into the data symbol transmitted by \emph{node A}, so that $\mean\{\ptr\} = 1$ (see \eqref{eq:trainpower} below), where the expectation is taken with respect to the distribution of $\hat \singval$ in \eqref{eq:tw92replicate}. Now, in the proof of Lemma \ref{lem:ptrcontrol}, using the probability density function (pdf) of $\norm{\hat{\channel}}_2$ in place of the pdf of $\singval$, and noting that the effective noise variance $\doteq 1/\barp$, we get $\kappone \doteq \frac{1}{\barp^{g_m/\alpha}}$ and the constraint $0 \le l \le r$ to satisfy $\mean\{\ptr\} = 1$ at high SNR.
In the next subsection, we explain the third round of training that alleviates the mismatch in the knowledge of the data transmit power.
\subsection{\textbf{Phase III (Power-Controlled Forward Link Training)}} \label{sec:3waytrain}
In this phase, \emph{node A} transmits the training sequence:
$\xAtraintwo = \sqrt{\barp \LAttwo} \sqrt{\ptr}$,
where $\LAttwo$ is the training duration. The corresponding received training signal at \emph{node B} is given by,
\beq \label{eq:trainpower}
\yBtrainvectwo =  \sqrt{\barp \LAttwo} \sqrt{\ptr}  \channel + \noisevecBttwo,
\eeq
where $\noisevecBttwo  \in \complex^{r\times 1}$ is the AWGN with $\standgauss$ entries. The goal at \emph{node B} is to estimate the composite channel $\ptrtrain \triangleq \sqrt{\ptr}  \channel$. Dividing \eqref{eq:trainpower} by $\sqrt{\barp \LAttwo}$, we get
\beq \label{eq:rcdpowertrain}
\yBtrainvectildetwo \triangleq \frac{\yBtrainvectwo}{\sqrt{\barp \LAttwo}} = \ptrtrain + \frac{\noisevecBttwo}{\sqrt{\barp \LAttwo}}.
\eeq
From \eqref{eq:rcdpowertrain}, \emph{node B} computes an MMSE estimate of $\ptrtrain$, denoted by ${\ptrtrainhat}$. Let ${\ptrtraintilde} \triangleq \ptrtrain -{\ptrtrainhat}$.
Although a closed form expression for $\ptrtrainhat$ is hard to find, the error $\ptrtraintilde$
in the MMSE estimate has the following interesting property, which
facilitates the calculation of the outage probability in Sec. \ref{sec:3waydmt}. An analogous result has been shown in  \cite{guo2008estimation} for the scalar case. 
\begin{lem} \label{lem:mmseprop}
$\mean{\norm{\ptrtraintilde}_2^{2z}} \preceq \frac{1}{\barp^{z}}$ for every $z>0$.
\end{lem}
\emph{Proof:} See Appendix \ref{app:mmseprop}.~$\blacksquare$

\subsection{\textbf{Phase IV (Data Transmission)}} \label{sec:3waydata}
Using $\ptr$, \emph{node A} sends the data signal $x = \sqrt{\barp \ptr}\xAdata$, where $\xAdata$ is distributed as $\mathcal{CN}(0,1)$ and is independent of $\ptr$. Note that $\mean{\abs{x}^2} = \barp$ by construction, where the expectation is taken with respect to both $\hat \singval$ and $x_{A,d}$. The corresponding signal received at \emph{node B} is
\beqa \label{eq:twowaydata}
\yBdatavec &=&  \sqrt{\barp \ptr} \channel \xAdata + \noisevecBd\\
&=& \sqrt{\barp} \ptrtrainhat \xAdata +  \sqrt{\barp} \ptrtraintilde \xAdata + \noisevecBd.
\eeqa
Since $\ptrtrainhat$ is an MMSE estimate, using the worst case noise theorem  \cite{ref3}, we have the following lower bound on the mutual information, $I(\xAdata;\yBdatavec|\ptrtrainhat) \geq C_{AB}$, where
\beq \label{eq:twowayinfo}
C_{AB} \triangleq \alpha \log\left(1 + \frac{\barp \norm{\ptrtrainhat}_2^2}{ \frac{\barp}{r} {\mean{[\norm{\ptrtraintilde}_2^2| \yBtrainvectildetwo]}} + 1}\right),
\eeq
and $\alpha \triangleq \frac{\lcoh - \LB - \LAtone - \LAttwo}{\lcoh}$ is the fractional data transmit duration after accounting for the time overheads in all three training phases. 

\subsection{DMT Analysis With Three-Way Training} \label{sec:3waydmt}
\begin{thm} \label{thm:dmttwoway}
For a SIMO system with $r$ receive antennas and three phases of training and the data transmission phase as described in Table \ref{twoway}, an achievable DMT is given by
\beq \label{eq:twowaydmt}
d(g_m) = r\left(\min\{l,s\} + 1 -\frac{g_m}{\alpha}\right),
\eeq
where $0 \leq l \leq r$, $1 \le s <r$, $0 \leq g_m < {\alpha}$, and $\alpha \triangleq \frac{\lcoh - \LB - \LAtone - \LAttwo}{\lcoh}$.
\end{thm}
\emph{Proof:} See Appendix \ref{app:dmttwoway}.~$\blacksquare$

\textbf{Remark}: The above three way training scheme can be generalized to $k$ training rounds to improve the diversity order, as in \cite{AggarwalSabharwalTITJune2011, ZhangGongLetaiefTWCJuly11}. However, this is mathematically cumbersome and out of the scope of our work. 

\section{Discussion} \label{sec:discussion}
\begin{figure}[t]
  \begin{center}
  \setxysize
    \includegraphics[width=3.0in]{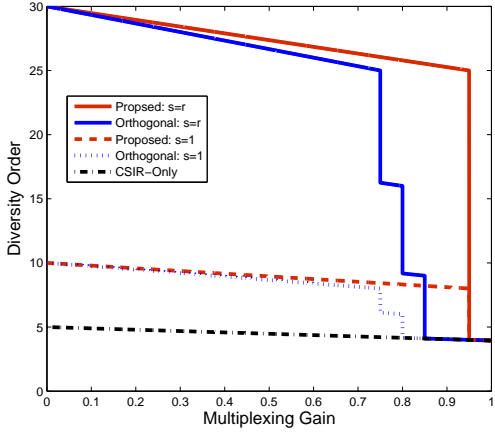}
  \caption{The achievable DMT with the training and power control scheme proposed in Sec. \ref{sec:syspftcsir}, compared with the performance with orthogonal RCT and the data power control proposed in \cite{StegerSabharwalTWCDec2008, ZhangGongLetaiefTWCJuly11} (and appropriately accounting for the training duration overhead and switching off antennas to achieve higher values of $g_m$). The plot corresponds to a SIMO system with $r=5$ antennas, with coherence time $\lcoh = 20$ symbols, and reverse training duration of $\LB = 1$ symbol.} \label{fig:DMT1}
  \end{center}
  \end{figure}

Recall that with perfect CSIR and imperfect CSIT, with $l \geq s+1$, and for a genie aided channel, it was shown in Theorem~\ref{thm:thmdmt1} that the following DMT is achievable:
\beq \label{eq:dmtrepeat}
d(g_m) = r \left[s+1-\left(\dfrac{g_m \lcoh}{\lcoh- \LB}\right)\right],
\eeq
where $1 \le s < r$, $0 \leq g_m \leq \dfrac{\lcoh- \LB}{\lcoh}$. In contrast, for the same genie aided channel, it was shown in \cite{StegerKhoshnevisSabharwalAazhangISIT2007} that a diversity order of
\beq \label{eq:SABHAR}
d_{s}(g_m) = r \left[2-\left(\dfrac{g_m \lcoh}{\lcoh-r \LB}\right)\right],~0 \leq g_m \leq \dfrac{\lcoh-r \LB}{\lcoh}
\eeq
is achievable using orthogonal reverse channel training. Note that $d_s(g_m)$ saturates as $r$ gets large, as opposed to \eqref{eq:dmtrepeat}, which is monotonically increasing in $r$. In order to achieve a $g_m > \frac{\lcoh-r \LB}{\lcoh}$,  in \cite{StegerSabharwalTWCDec2008}, the authors suggest turning off one receive antenna at a time to reduce the training burden until $r=2$. For example, turning off one antenna, $g_m \in \left[\frac{\lcoh-r \LB}{\lcoh}, \frac{\lcoh-(r-1) \LB}{\lcoh}\right]$ is achievable at a reduced diversity order of $d_s(g_m) = (r-1) \left[2-\left(\frac{g_m \lcoh}{\lcoh-(r-1) \LB}\right)\right]$. This is in contrast to our result, which can accommodate a larger multiplexing gain, $g_m \leq \frac{\lcoh- \LB}{\lcoh}$ irrespective of $r$, while simultaneously achieving a higher diversity order at each $g_m$. We note that for a SIMO channel, a diversity order of $r(r+1 - g_m)$ for $0 \le g_m < 1$ was obtained in \cite{KimCaireTITApr2009, ZhangGongLetaiefTWCJuly11}, using channel-independent training, and without accounting for the training duration overhead. This corresponds to taking $\lcoh \rightarrow \infty$ in \eqref{eq:dmtrepeat}. The performance of the proposed scheme is schematically contrasted with orthogonal RCT in Fig. \ref{fig:DMT1} for a SIMO system with $r=5$, $\lcoh = 20$, and $\LB = 1$ symbol.
The advantage of the proposed scheme at higher values of the multiplexing gain is clear from the plot.  
The proposed training scheme thus results in a \emph{factor $r$-reduction} in the training duration, which, along with the proposed data power control scheme, translates to an increase in the range of achievable multiplexing gains, while simultaneously offering a better diversity order compared to orthogonal RCT schemes.

Comparing Theorems \ref{thm:thmdmt1} and \ref{thm:dmttwoway}, we see that the DMT performance of a genie aided receiver with perfect CSIR is an upper bound on the performance of the system with imperfect CSIR and CSIT, as expected. Also, the performance of the two systems is similar, except that in the latter case, the factor $\alpha$ captures the loss in data transmission time due to all three training phases. Similar observations as the above regarding the improvement in DMT can be made for the three way training scheme compared to orthogonal RCT schemes.

\section{Simulation Results} \label{sec:sims}
We now briefly present Monte-Carlo simulation results to illustrate the outage probability performance of our proposed RCT and forward-link data power control schemes.  We consider a Rayleigh fading channel with three receive antennas.  We calculate the outage probability by averaging over $10^8$ \emph{i.i.d.} channel and training noise instantiations. We set the channel coherence time and reverse training duration as $\lcoh = 40$ and $\LB = 1$, respectively.
 \begin{figure}[t]
  \begin{center}
  \setxysize
  \includegraphics[width=3.0in]{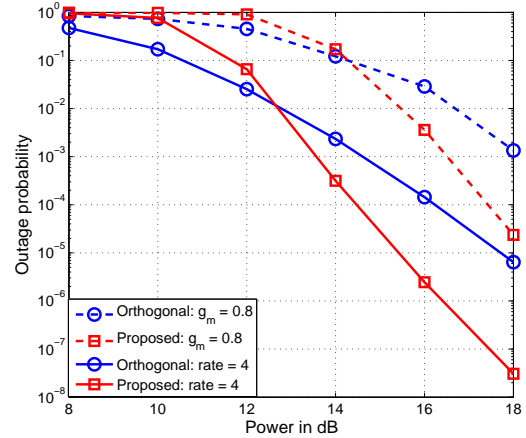}
  \caption{Outage probability versus the average data power $\barp$  for the fixed-power training scheme proposed in Sec. \ref{sec:syspftcsir}, with the data power control scheme given by \eqref{eq:DATRULE} with $s=1$. Here, $r=3, \lcoh = 40$ and $\LB = 1$. With $g_m=0.8$, the target data rate was set as $R_{\barp} = 4 + g_m \log \barp$ to facilitate the comparison of the curves. } \label{fig:RtwoR}
  \end{center}
  \end{figure}
     \begin{figure}[t]
   \begin{center}
   \setxysize
   \includegraphics[width=3.0in]{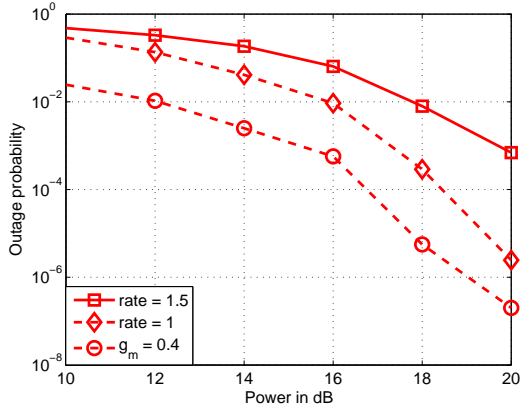}
   \caption{Outage probability versus the average data power $\barp$  for the fixed-power training scheme proposed in Sec. \ref{sec:syspftcsir}, with the data power control scheme given by \eqref{eq:DATRULE} with $s=r$.  Here, $r=3, \lcoh = 40$ and $\LB = 1$.} \label{fig:RintoRplus1}
   \end{center}
   \end{figure}
Figures \ref{fig:RtwoR} and \ref{fig:RintoRplus1} show the outage probability of the proposed fixed-power training scheme and the data power control scheme in \eqref{eq:DATRULE} with $s=1$ and $s=r=3$, respectively,  as a function of $\barp$, with $g_m=0$ and $R_{\barp}=4$ bits/channel use ($1$ and $1.5$ bits/channel use in case of Fig. \ref{fig:RintoRplus1}), and with $g_m = 0.8$. 
Although the slopes of the curves do not match with the theoretical diversity order because the latter requires infinite SNR, the improved performance of the proposed schemes
is clear from the graphs. Also, in Fig. \ref{fig:RtwoR}, since the proposed scheme uses only $\LB = 1$ training symbol while the orthogonal RCT scheme uses $r\LB = 3$ training symbols, the former shows a higher outage than the latter at lower SNRs.
Note that, we have not plotted the outage performance of the three-way training scheme in Sec.~\ref{sec:twoway}. This is because the outage probability is hard to compute, since a closed-form expression for $\hat{\mathbf{p}}_c$ is not available.

\section{Conclusions} \label{sec:con}
This paper proposed reverse training and data power control schemes for a TDD-SIMO system with perfect/imperfect CSIR and investigated its DMT performance. It was shown that a diversity order of $d(g_m) = r\left(s+1 - \frac{g_m}{\alpha}\right)$ is achievable for $l \geq s+1$, $1 \le s < r$ and $0 \le g_m < \alpha$, where $\alpha$ represents the fractional data transmit duration. In contrast to channel agnostic orthogonal training schemes, the diversity order was shown to increase monotonically with $r$ at nonzero multiplexing gain, which is a significant improvement. 
The DMT analysis was extended to a more practical situation where the training is done in both directions. In this case also, it was shown that the DMT performance can improve quadratically with the number of receive antennas, and nearly the same DMT can be achieved as that with perfect CSIR and a genie-aided receiver. In terms of system design for reciprocal SIMO systems, the key messages from this work are that it is important to (a) exploit the CSI at the receiver in designing the RCT and (b) use a modified channel-inversion type power control scheme that transmits data at some non-zero power even when the estimated singular value at the transmitter is poor. For fast varying channels, these ingredients can lead to
a significant advantage in DMT performance, which, at finite SNR, can translate to a large improvement in outage probability performance compared to orthogonal training schemes.
Future work could extend the DMT analysis to a time-selective block fading reciprocal channel, where the channel is correlated within a block~\cite{liang2004capacity}.

\appendix \label{sec:append}
\setcounter{subsubsection}{0}
\subsection{Useful Lemmas}
\begin{lem} \label{lem:PIone}
If the random variable $\sigma^2$ is a chi-square distributed with $2r$ degrees of freedom, then $\Pr\{\sigma^2 < z\} \leq \frac{z^r}{r!}$, $z \geq 0$.
\end{lem}
\emph{Proof}: The result follows from
\beqa
\Pr\{\sigma^2 < z\}  &=& \frac{1}{{(r-1)!}}\int_{0}^{z}{e^{-x} x^{r-1} dx}\\
&\leq& \frac{1}{{(r-1)!}}\int_{0}^{z}{x^{r-1} dx}
= \frac{z^r}{r!}.~\blacksquare
\eeqa

\begin{lem} \label{lem:sigmaupper}
For the system in \eqref{eq:tw92}, $\abs{\hat \singval} \leq \hat{\singval}_U,$
where $\hat{\singval}_U^2 \triangleq \left(\singval + \abs{\noisescalarAtbar}\right)^2$, with ${\noisescalarAtbar} \triangleq {\frac{\Re\{\noisescalarAt\}}{\sqrt{\barp \LB}}}$.
\end{lem}
\emph{Proof}: We upper bound the absolute value of \eqref{eq:tw92} as follows:
\beq
\abs{\hat \singval}
\stackrel{(a)}{\leq} {\singval} \abs{\Re\{\singvec^H \hat{\singvec}\}} + \abs{\frac{\Re\{{\noisescalarAt}\}}{\sqrt{\barp \LB}}}
\stackrel{(b)}{\leq} {\singval}  + \abs{\noisescalarAtbar},
\eeq
where \emph{(a)} follows from the triangle inequality and \emph{(b)} follows since $\abs{\Re\{\singvec^H \hat{\singvec}\}} \leq 1$.~$\blacksquare$

\subsection{Proof of Lemma \ref{lem:ptrcontrol}} \label{app:ptrcontrol}
Consider the following constraint on the data power
\beq \label{eq:PCONTR}
\mathbb{E}[\ptr] = \int_{- \infty}^\infty{\ptr f_{\hat{\singval}}(\hat{\singval};\bar{P})d\hat{\singval}} = \bar{P},
\eeq
where $f_{\hat{\singval}}(\hat{\singval};\bar{P})$ is the pdf
of $\hat{\singval}$.
Substituting \eqref{eq:DATRULE} in \eqref{eq:PCONTR}, we get
\beq \label{eq:secPC1}
\mean[\ptr] =\kappone \left[\exp\left(\frac{\lcoh R_{\barp}}{\lcoh-\LB}\right)-1\right] F(\bar{P})+ { I_{\barp}},
\eeq
where $R_{\barp}$ is the target data rate and the data transmit power is $\barp$,
\beq \label{eq:secPC2_case2}
F(\bar{P}) \triangleq  \int_{\theta_{\barp}}^{\infty} {\frac{1}{x^{2s}} f_{\hat{\singval}}(x;\bar{P})} dx ~~ \text{and}~~ I_{\barp} \triangleq  \bar{P}^l \int_{-\infty}^{\theta_{\bar{P}}} {f_{\hat{\singval}}(x;\bar{P})} dx.
\eeq
The proof is complete by choosing
\beq \label{eq:kapp}
\kappone = \frac{1}{\left(\exp\left(\frac{\lcoh R_{\barp}}{\lcoh - \LB}\right)-1\right)F(\barp)}(\bar{P} - I_{\barp}),
\eeq
and showing that $I_{\barp} < \barp$ and that $F(\barp)$ is bounded for large $\barp$ when $0 \leq l \leq r+1$ and $n = 1/2$.
From \eqref{eq:secPC2_case2}, $I_{\barp} = \bar{P^l} \Pr\{\singval + \noisescalarAtbar < \theta_{\bar P}\}$ can be bounded as,
\beqa \label{eq:Secapa1}
I_{{\bar{P}}}
&\stackrel{(a)}{\leq}& \frac{\bar{P^l}}{r!} \mean{(\theta_{\barp} - \noisescalarAtbar)^{2r}}
\stackrel{(b)}{=} \frac{\bar{P^l}}{r!} \mean{\sum_{j=0}^{r}{\theta_{\barp}^{2(r-j)} {2r \choose  2j} \noisescalarAtbar^{2j}}}\nonumber \\
&\stackrel{(c)}{\doteq}& \barp^l \max_{j \in \{0,1,\ldots,r\}} \frac{1}{\barp^{2(r-j)n + j}}\label{eq:doteqL} \stackrel{(d)}{\doteq} \frac{1}{\bar P^{r-l}},
\label{eq:Secapa1d}
\eeqa
where \emph{(a)} follows from Lemma \ref{lem:PIone} above, and the expectation is with respect to the distribution of $\noisescalarAtbar$, \emph{(b)} follows from the binomial expansion and the fact that $\mean{\noisescalarAtbar^{i}}= 0$ when $i$ is odd, \emph{(c)} follows from $\theta_{\barp} \doteq \frac{1}{\barp^n}$ and $\mean{\noisescalarAtbar^{2j}} \doteq \frac{1}{\barp^{j}}$, and \emph{(d)} follows by substituting $n = 1/2$ in the left hand side. From \eqref{eq:Secapa1d}, clearly, $I_{\barp} < \barp$ for large $\barp$ if $l < r+1$ and $n = 1/2$. When $l = r+1$ and $n = 1/2$, we have $I_{\barp} \preceq \barp$, and therefore we can ensure that $I_{\barp} < \barp$ for large $\barp$ by scaling $I_{\barp}$ by an appropriately chosen constant scaling factor.

Next, we show that $F(\barp)$ is bounded.  Note that
\beqa
F(\bar{P}) \hspace{-2.8mm}&=& \hspace{-2.8mm}\int_{\theta_{\barp}}^{1} {\frac{1}{x^{2s}} f_{\hat{\singval}}(x;\bar{P})} dx
+\int_{1}^{\infty} {\frac{1}{x^{2s}} f_{\hat{\singval}}(x;\bar{P})} dx. \label{eq:secPCADD1}
\eeqa
Now, it is sufficient to show that the first integral in \eqref{eq:secPCADD1} is bounded, since the second integral is clearly $<1$.
To this end, we need the distribution  of $\hat{\singval}$, i.e., $\Pr\left(\singval + \noisescalarAtbar \leq x\right)$, where $\noisescalarAtbar \thicksim \mathcal{N}(0,\sigma_{var}^2)$, and $\sigma_{var}^2 \triangleq \frac{1}{2\bar{P}\LB}$. Consider
\beqa \label{eq:secaPC2}
G(x) &\triangleq& \hspace{-2.5mm}\Pr\left(\singval + \noisescalarAtbar \leq x \right)\\
&=& \hspace{-2.5mm} \int_{0}^\infty{\hspace{-2.5mm} f_\singval(y)\int_{-\infty}^{x-y}\frac{1}{\sqrt{2 \pi }\sigma_{var}}e^{-z^2/{2 \sigma_{var}^2}}dz dy},
\eeqa
where $f_{\singval}({y})$ is the pdf of $\singval$, which is chi distributed with $2r$ degrees of freedom. Taking the derivative of \eqref{eq:secaPC2} with respect to $x$, we get
\begin{align} \label{eq:secaPC3}
\frac{\partial G(x)}{\partial x}
&=& \frac{J}{\sqrt{2 \pi}\sigma_{var}}\int_{0}^\infty{y^{2r-1} e^{-\frac{y^2}{2}} e^{-\frac{(x-y)^2}{{2 \sigma_{var}^2}}}dy}\\
&=& \frac{J e^{-\beta_3}}{\sqrt{2 \pi}\sigma_{var}}\int_{0}^\infty{y^{2r-1} e^{\left\{-\frac{(y-\beta_1)^2}{{2\beta_2}}\right\}}dy},
\end{align}
where $J$ is the constant term in the standard chi pdf,
$\beta_1 \triangleq \frac{x}{1+\sigma_{var}^2}$, $\beta_2 \triangleq \frac{\sigma_{var}^2}{1+\sigma_{var}^2} \doteq \frac{1}{\barp}$ and
$\beta_3 \triangleq \beta_2 x^2/(2\sigma_{var}^2)$. Let $t = \frac{y-\beta_1}{\sqrt{\beta_2}}$ and using the binomial expansion, it can be shown that
\ber \label{eq:secaPC5}
\frac{\partial G(x)}{\partial x}
&=&  \frac{J \exp(-\beta_3)}{\sqrt{2 \pi}\sigma_{var}} \sum_{j=0}^{2r-1}{2r-1 \choose j}(\sqrt{\beta_2})^{2r-j}\nonumber \\ &\times&\frac{x^j}{(1+\sigma_{var}^2)^j} {\int_{-\beta_1/\sqrt{\beta_2}}^\infty{t^{2r-1-j} e^{-\frac{t^2}{2}}dt}}.
\eer

Now, using  $\exp(-\beta_3) \leq 1$, we can upper bound the first term in \eqref{eq:secPCADD1} as
\beqa
\int_{\theta_{\barp}}^{1}{\frac{1}{x^{2s}}\frac{\partial G(x)}{\partial x} dx} &\leq&  \frac{J}{\sqrt{2 \pi}\sigma_{var}} \sum_{j=0}^{2r-1}{2r-1 \choose j}C_j \nonumber\\
& \times& \frac{(\sqrt{\beta_2})^{2r-j}}{(1+\sigma_{var}^2)^j} \int_{\theta_{\barp}}^1{x^{j-2s}dx}, \label{eq:secaPC6case2}
\eeqa
where $s<r$, and $C_j \doteq 1$ is some constant that does not scale with $\barp$. Now, the behavior of the terms above with $\barp$ is governed by
\beq
\frac{\beta_2^{r-j/2}}{\sigma_{var}} \int_{\theta_{\barp}}^1{x^{j-2s}dx} \doteq \frac{1}{j - 2s + 1}\left[\frac{1}{\barp^{a_1}} - \frac{1}{\barp^{a_2}}\right],
\eeq
where $a_1 \triangleq r-j/2-1/2$, and $a_2 \triangleq (-2s + j + 1)n+r-j/2-1/2$.
The exponent corresponding to the first term above is $r - j/2 - 1/2 \geq 0$ for all $0 \leq j \leq 2r -1$. 
Also, when $n=1/2$, the exponent corresponding to the second term above is $r-s >0$ for all $0 \le j \le 2r-1$, and hence the integral is bounded for $1 \le s < r$.

Finally, let $R_{\barp} = g_m \log(\barp)$. Since $I_{\barp} < \barp$ and $F(\barp)$ are bounded when $0 \le l \le r+1$, using $\left(\exp\left(\frac{\lcoh R_{\barp}}{\lcoh - \LB}\right)-1\right) \doteq \barp^{\frac{g_m}{\alpha}}$ in \eqref{eq:kapp}, we get $\kappone \doteq \frac{1}{\barp^{\frac{g_m}{\alpha}-1}}$, where  $\alpha \triangleq \frac{\lcoh - \LB}{\lcoh}$. This completes the proof of Lemma~\ref{lem:ptrcontrol}. $\blacksquare$

\subsection{Proof of Theorem \ref{thm:thmdmt1}} \label{app:thmdmt1}
Using the power control in \eqref{eq:DATRULE}, the outage probability in \eqref{eq:OUT} can be written as
\beqa \label{eq:outproof1}
P_{out} &=& \Pr_{\{\hat{\singval} \leq \theta_{\barp}\}}\left\{\alpha \log(1+\bar{P}^l\sigma^2)<R_{\barp} \right\}\\
&+& \Pr_{\{\hat{\singval} > \theta_{\barp}\}}\left\{\alpha \log(1+\kappone \Phi(\hat{\singval}^{2s})\sigma^2) < R_{\barp}\right\}\\
&\leq& \Pi_1 + \Pi_2,\label{eq:outproof1a}
\eeqa
where $\Pi_1 \triangleq \Pr\left\{\alpha \log(1+\bar{P}^l\sigma^2)<R_{\barp} \right\}$, and $\Pi_2 \triangleq \Pr\left\{\alpha \log(1+\kappone \Phi(\hat{\singval}^{2s})\sigma^2) < R_{\barp}\right\}$.
In the above, we have used $\Pr_{\{A\}}\{\cdot\}$ to mean $\Pr\{\cdot \bigcap \{A\}\}$.
Using $R_{\barp} = g_m \log{\barp}$, we have $\Pi_1 = \Pr\left\{\sigma^2 < \frac{1}{\barp^{l-\frac{g_m}{\alpha}}}\right\}$ for large $\barp$ and $0 \leq l \leq r+1$  from Lemma~\ref{lem:ptrcontrol}. From  Lemma~\ref{lem:PIone} in Appendix~\ref{sec:append}, we have,
$\Pi_1 \preceq \frac{1}{\barp^{\left(l-\frac{g_m}{\alpha}\right)r}}$.
Next, substituting for $\Phi(\hat{\singval}^{2s})$ from \eqref{eq:optpalloc}, $\Pi_2$ can be written as,
$\Pi_2 = \Pr\left\{\singval^2 < {{\hat \singval}^{2s}}/{\kappone} \right\}.$
Using $\hat{\singval}^{2} \leq \hat{\singval}^2_U \triangleq (\singval +|\noisescalarAtbar|)^2$ from Lemma~\ref{lem:sigmaupper} in Appendix~\ref{sec:append} with $\hat{\singval}^2 = \singval^2$, we get
\ber \label{eq:Pi1_1_case2}
\Pi_2 &\leq& \Pr\left\{\singval^2 < \frac{1}{\kappone}(\singval + |\noisescalarAtbar|)^{2s}\right\}\\
&\leq& \Pr\left\{\singval^2 < \frac{(2 \singval)^{2s} }{\kappone}\bigcap \singval^2 > |\noisescalarAtbar|^2\right\}\nonumber \\
&+& \Pr\left\{\singval^2 < \frac{(2 \abs{\noisescalarAtbar})^{2s}}{\kappone} \bigcap \singval^2 \leq |\noisescalarAtbar|^2\right\}. \label{eq:Pi1_2_case2}
\eer
It is straightforward to show that provided $\kappone$ is strictly increasing with $\barp$, the first term in the above goes to zero exponentially with $\barp$ for $1 \le s < r$. This implies that $g_m < \alpha$, since $\kappone \doteq \barp^{\left(1-\frac{g_m}{\alpha}\right)}$ from Lemma~\ref{lem:ptrcontrol}.
The second term in \eqref{eq:Pi1_2_case2} is upper-bounded as
\beqa
\Pr\left\{\sigma^2 < \frac{|\noisescalarAtbar|^{2s} 2^{2s}}{\kappone}\right\} &\stackrel{(a)}{\leq}& \frac{2^{2sr}\mean{|\noisescalarAtbar|^{2sr}}}{\kappone^r r!}\\
&\stackrel{(b)}{\doteq}& \frac{1}{\barp^{r\left(s+1-\frac{g_m}{\alpha}\right)}},
\eeqa
where \emph{(a)} follows from  Lemma \ref{lem:PIone} in Appendix \ref{sec:append}, and the $\doteq$ in \emph{(b)} uses the fact that $\kappone \doteq \barp^{\left(1-\frac{g_m}{\alpha}\right)}$ and $ \mean{|\noisescalarAtbar|^{2sr}} \doteq 1/\barp^{sr}$. Hence, we have
$\Pr\left\{\singval^2 < \frac{|\noisescalarAtbar|^{2s} 2^{2s}}{\kappone}\right\} \preceq \frac{1}{\barp^{r\left(s + 1 -\frac{g_m}{\alpha}\right)}},$ 
which implies
$\Pi_2 \preceq \frac{1}{\barp^{r\left(s+1 - \frac{g_m}{\alpha}\right)}}$.
Using this and $\Pi_1 \preceq \frac{1}{\barp^{r \left(l-\frac{g_m}{\alpha}\right)}}$  in \eqref{eq:outproof1a}, we have
\beqa
P_{out} &\preceq& \max \left(\frac{1}{\barp^{r\left(l - \frac{g_m}{\alpha}\right)}},\frac{1}{\barp^{r\left(s+1 - \frac{g_m}{\alpha}\right)}}\right)\\ 
&=& \frac{1}{\barp^{r\left(\min\{l,s+1\} - \frac{g_m}{\alpha}\right)}}, \label{eq:DMTFINAL}
\eeqa
for $0 \leq l \leq r+1$, $1 \le s < r$ and $0 \leq g_m < \alpha$. This ends the proof of Theorem~\ref{thm:thmdmt1}.~$\blacksquare$

\subsection{Proof of Lemma \ref{lem:mmseprop}} \label{app:mmseprop}
Note that $\ptrtraintilde$ can be written as
\beqa
\ptrtraintilde 
&=& \ptrtrain - \yBtrainvectildetwo - \mean{\{\ptrtrain - \yBtrainvectildetwo|\yBtrainvectildetwo\}}\\
&=& \frac{1}{\sqrt{\barp \LAttwo}} \left[\mean{\{\noisevecBttwo|\yBtrainvectildetwo\}} - \noisevecBttwo\right].
\eeqa
Now,
\beqa
\mean{\norm{\ptrtraintilde}_2^{2z}} &=&  \frac{1}{{\barp^{z} \LAttwo^{z}}} \mean_{\noisevecBttwo,\yBtrainvectildetwo} \left\{ \mathcal{A}\right\}\label{eq:msebound1}\\
&\stackrel{(a)}{\leq}& \frac{1}{{\barp^{z} \LAttwo^{z}}} \left[\mean_{\yBtrainvectildetwo} \left\{2^{2z}\norm{\mean{\{\noisevecBttwo|\yBtrainvectildetwo\}}}_2^{2z}\right\}\right.\nonumber \\
&~&~+ \left.2^{2z}\mean_{\noisevecBttwo}\left\{ \norm{\noisevecBttwo}_2^{2z}\right\}\right]\\
&\stackrel{(b)}{\leq}& \frac{2^{2z+1}}{{\barp^{z} \LAttwo^{z}}}  \mean \norm{\noisevecBttwo}_2^{2z} \doteq  \frac{1}{{\barp^{z}}},\label{eq:msebound2}
\eeqa
where $\mathcal{A} \triangleq \norm{ \mean{\{\noisevecBttwo|\yBtrainvectildetwo\}} - \noisevecBttwo}_2^{2z}$. In the above, \emph{(a)} follows from the triangle inequality and using $(a+b)^n \leq (2a)^n +(2b)^n$ for even $n > 0$, and \emph{(b)} follows from the Jensen's inequality and the fact that $\mean \norm{\noisevecBttwo}_2^{2z} < \infty$ as $\barp \rightarrow \infty$. The subscripts on the expectation in the above denote the random variables over which the expectation is taken; and $\mean\{X|y\}$ denotes the expectation of the random variable $X$ conditioned on the instantiation $Y=y$. This completes the proof.~$\blacksquare$

\subsection{Proof of Theorem \ref{thm:dmttwoway}} \label{app:dmttwoway}
Using the capacity lower bound in \eqref{eq:twowayinfo}, the outage probability can be upper bounded as
\beq
P_{out} \le \Pr\left\{C_{AB} < R_{\barp}\right\},
\eeq
where $R_{\barp} \triangleq g_m \log \barp$ is the target data rate. We choose $\eta < 1$, and arbitrarily close to $1$. We split the event in the expression for $P_{out}$ as
\beqa
P_{out} &\leq& \Pr\left\{C_{AB} < R_{\barp} \cap \mean{[\norm{\ptrtraintilde}_2^2| \yBtrainvectildetwo]} \leq \frac{1}{\barp^{\eta}}\right\}\\
&~& + \Pr\left\{C_{AB} < R_{\barp} \cap \mean{[\norm{\ptrtraintilde}_2^2| \yBtrainvectildetwo]} > \frac{1}{\barp^\eta}\right\} \nonumber\\
&\stackrel{(a)}{\leq}& \Pr\left\{\alpha \log\left(1 + \frac{\barp \norm{\ptrtrainhat}_2^2}{ \frac{\barp^{(1-\eta)}}{r} + 1}\right) < R_{\barp}\right\} \nonumber \\
&~& + \Pr\left\{\mean{[\norm{\ptrtraintilde}_2^2| \yBtrainvectildetwo]} > \frac{1}{\barp^{\eta}} \right\}, \label{eq:poutexpsecondterm}
\eeqa
where \emph{(a)} follows by substituting $1/\barp^\eta$ in place of $\mean{[\norm{\ptrtraintilde}_2^2| \yBtrainvectildetwo]}$ in the first term, and removing one of the events in the intersection.
Define $\bar{R}_{\barp} \triangleq \frac{(\exp\{{R_{\barp}}/{\alpha}\}-1)}{\barp} \left(\frac{\barp^{(1-\eta)}}{r} + 1\right)$, and note that $\bar{R}_{\barp} \doteq \frac{1}{\bar{P}^{\lb\eta-\frac{g_m}{\alpha}\rb}}$.  Then, the first term in \eqref{eq:poutexpsecondterm} can be written as:
\ber
\Pr\left\{{\norm{\ptrtrainhat}_2^2} < {\bar{R}_{\barp}}\right\} 
\hspace{-3mm}&\stackrel{(a)}{\leq}& \hspace{-3mm}\Pr\left\{|\norm{\ptrtrain}_2-\norm{\ptrtraintilde}_2| < \sqrt{\bar{R}_{\barp}}\right\}\label{eq:twin5d}\\
\hspace{-3mm}&\leq& \hspace{-3mm} \Pr\left\{E_1 \bigcap E_2\right\} + \Pr\left\{E_1 \bigcap E_2^c\right\} \nonumber \\
\hspace{-3mm}&\leq& \hspace{-3mm}\Pr\left\{\norm{\ptrtraintilde}_2 > \sqrt{\bar{R}_{\barp}}\right\}\nonumber \\
&~&~~~~+ \Pr\left\{\norm{\ptrtrain}_2^2 < 4 {\bar{R}_{\barp}}\right\}, \label{eq:twin5g}
\eer
where $E_1 \triangleq \{\norm{\ptrtrain}_2 < \norm{\ptrtraintilde}_2  + \sqrt{\bar{R}_{\barp}}\}$ and $E_2 \triangleq \{\norm{\ptrtraintilde}_2 > \sqrt{\bar{R}_{\barp}}\}$.
In the above, \emph{(a)} follows from the reverse triangle inequality, and the last two inequalities follow by ignoring one of the events in the intersection. The first term in \eqref{eq:twin5g} can be written as
\ber \label{eq:LSMMSE}
\Pr\left\{\norm{\ptrtraintilde}_2^{2\delta} > {\bar{R}_{\barp}}^\delta\right\}
\stackrel{(a)}{\leq} \frac{\mathbb{E}\norm{\ptrtraintilde}_2^{2\delta}}{\bar{R}_{\barp}^\delta}
\stackrel{(b)}{\preceq} \frac{1}{\bar P^{\delta}} \frac{1}{\bar{P}^{{\left(\frac{g_m}{\alpha}-\eta\right)\delta}
}},
\eer
where \emph{(a)} follows from the Markov inequality and \emph{(b)} follows from Lemma \ref{lem:mmseprop}. Letting $\delta = r \frac{1}{\frac{g_m}{\alpha} - \eta + 1} \left(s + 1-\frac{g_m}{\alpha}\right)>0$, we have
\begin{equation} \label{eq:goodbad1}
\Pr\left\{\norm{\ptrtraintilde}_2 > \sqrt{\bar{R}_{\barp}}\right\} \preceq \frac{1}{\bar{P}^{r\left(s+1-\frac{g_m}{\alpha}\right)}}, ~1 \le s < r.
\end{equation}
In order to solve for the {second} term in \eqref{eq:twin5g}, we need to handle two cases of the singular value estimate at \emph{node A} separately; the good estimated channel case $g \triangleq \{\hat{\singval} \geq {\theta}_{\barp}\}$ and the bad estimated channel case $b \triangleq \{\hat{\singval} < {\theta}_{\barp}\}$.

\subsubsection{Good Estimated Channel $\{\hat{\singval} \geq {\theta}_{\barp}\}$}
When $\hat{\singval} \geq {\theta}_{\barp}$, substituting for $\ptrtrain \triangleq \sqrt{\ptr}  \channel$ and $\kappone \doteq \barp^{-\frac{g_m}{\alpha}}$, and defining $\hat{\singval}_U \triangleq \left(\singval + \abs{\noisescalarAtbar}\right)$ as the upper bound on $\hat{\singval}$ from Lemma \ref{lem:sigmaupper} in  Appendix \ref{sec:append},  the second term in \eqref{eq:twin5g} leads to:
\beqa
\Pr_{\{\hat{\singval} \geq {\theta}_{\barp}\}} \left\{E_3\right\}
&\stackrel{(a)} {\leq}& 
\Pr\left\{\singval^2 < {2^{2(s+1)} \singval^{2s}} \bar{R}_{\barp} \bigcap E_4\right\}\nonumber \\
&+& \Pr\left\{\singval^2 < {2^{2(s+1)} \abs{\noisescalarAtbar}^{2s}} \bar{R}_{\barp} \bigcap E_4^c\right\}\nonumber\\
&\leq&  \Pr\left\{\singval^{2(s-1)} > \frac{2^{-2(s+1)}}{ \bar{R}_{\barp}}\right\}\nonumber \\
&~& + \Pr\left\{\singval^2 < {2^{2(s+1)} \abs{\noisescalarAtbar}^{2s}} \bar{R}_{\barp} \right\}, \label{eq:firstterm_pout}
\eeqa
where $E_3 \triangleq \{\dfrac{\norm{\channel}_2^2}{\hat{\singval}_U^{2s}}<4 \bar{R}_{\barp}\}$, and $E_4 \triangleq \{\singval^2 > |\noisescalarAtbar|^2\}$. In the above, we have used $\Pr_{\{A\}}\{\cdot\}$ to mean $\Pr\{\cdot \bigcap \{A\}\}$, as before; and \emph{(a)} follows by ignoring the event  $g$. It can be shown that first term in \eqref{eq:firstterm_pout} decreases exponentially with $\barp^{\frac{\eta - g_m/\alpha}{s-1}}$, as follows:
\beqa
\Pr\left\{\mathcal{B}\right\}
&\stackrel{(a)}{\doteq}& \hspace{-3mm}  \int_{{1}/{\bar{R}_{\barp}^{1/(s-1)}}}^\infty e^{-x} x^{r-1} dx\\
&{\doteq}& \hspace{-3mm}{\exp\left\{-{1}/{\bar{R}_{\barp}^{1/(s-1)}}\right\}} \sum_{k=0}^{r-1} \dfrac{1}{(\bar{R}_{\barp}^{1/(s-1)})^{r-k-1}}\nonumber \\
&\stackrel{(b)}{\doteq}& \hspace{-3mm}e^{-\mathcal{Z}},
\eeqa
where $\mathcal{B} \triangleq \{\singval^{2(s-1)} > \frac{1}{2^{2(s+1)} \bar{R}_{\barp}}\}$, and $\mathcal{Z} \triangleq \barp^{\frac{\eta - g_m/\alpha}{s-1}}$. In the above, \emph{(a)} follows by ignoring the constant factors and substituting for the chi-square pdf of $\sigma^2$. Since $1/\bar{R}_{\barp} \doteq \barp^{(\eta - g_m/\alpha)}$ when $g_m < \eta \alpha$, and since the exponential term outside the summation dominates the polynomial terms inside the summation, we obtain \emph{(b)}. Note that the special case of $s=1$ is trivial,  since this corresponds to the probability that $\bar{R}_{\barp}$ exceeds a constant, which becomes $0$ for sufficiently large $\barp$.
The second term in \eqref{eq:firstterm_pout} becomes:
\beqa
\Pr\left\{\singval^2 < {2^{2(s+1)} \abs{\noisescalarAtbar}^{2s}} \bar{R}_{\barp} \right\} &\leq&
\frac{2^{2r(s+1)} \bar{R}_{\barp}^r \mean\{\abs{\noisescalarAtbar}^{2sr}\}}{r!}\nonumber \\
&\doteq& \frac{1}{\barp^{\left(\eta-\frac{g_m}{\alpha}\right)r }\barp^{rs}}\\
&=& \frac{1}{\barp^{r \left(s + \eta -g_m/\alpha \right)}}
\eeqa
for $ 0 \leq g_m < \eta \alpha$.
In the above, we have used Lemma \ref{lem:PIone} in Appendix \ref{sec:append} and  $\mean{\abs{\noisescalarAtbar}^{2s}} =\frac{1}{\barp^s}$.
Thus, in the good estimated channel case, we have
\beq
\Pr_{\{\hat{\singval} \geq {\theta}_{\barp}\}}\left\{\norm{\ptrtrain}_2^2 \leq 4{\bar{R}_{\barp}} \right\}\preceq \dfrac{1}{\bar{P}^{r\left(s + \eta-\frac{g_m}{\alpha}\right)}},~ 0 \leq g_m < \eta \alpha. \label{eq:good2}
\eeq

\subsubsection{Bad Estimated Channel $\{\hat{\singval} < \theta_{\barp}\}$}
Recall that when $\hat \singval < \theta_{\barp}$, the composite channel is given by $\ptrtrain = \sqrt{\bar{P}^l} \channel$. With this, the second term in \eqref{eq:twin5g} becomes
\beqa
\Pr_{\{\hat{\singval} < \theta_{\barp}\}}\left\{\norm{\ptrtrain}_2^2 < 4 \bar{R}_{\barp} \right\} &=& \Pr\left\{\norm{\channel}_2^2 < \dfrac{4 \bar{R}_{\barp}}{\bar P^{l}} \bigcap b\right\}\nonumber \\
&\leq& \Pr\left\{\singval^2 < \frac{4 \bar{R}_{\barp}}{\bar P^{l}}\right\}\\
&\doteq& \frac{1}{\bar P^{rl}} \frac{1}{\bar P^{r\left(-\frac{g_m}{\alpha}+\eta\right)}}\\ &\doteq& \frac{1}{\barp^{r\left(l + \eta - \frac{g_m}{\alpha}\right)}}, \label{eq:bad2}
\eeqa
where $0 \leq l \leq r$. This completes the analysis of the first term in \eqref{eq:poutexpsecondterm}.

Now, the second term in \eqref{eq:poutexpsecondterm} can bounded as:
\beqa
\Pr\left\{(\mean{[\norm{\ptrtraintilde}_2^2| \yBtrainvectildetwo]})^\zeta > \frac{1}{\barp^{\zeta\eta}} \right\} 
\hspace{-2.7mm}&\stackrel{(a)} {\leq}& \hspace{-2.7mm}\mean  (\mean{[\norm{\ptrtraintilde}_2^2| \yBtrainvectildetwo]})^\zeta \barp^{\zeta\eta}\nonumber \\
\hspace{-2.7mm}&\stackrel{(b)} {\leq}&  \hspace{-2.7mm}\mean({[\norm{\ptrtraintilde}_2^{2\zeta}]}) \barp^{\zeta\eta}\\
\hspace{-2.7mm}&\stackrel{(c)} {\preceq}& \hspace{-2.7mm}\frac{1}{\barp^{\zeta(1- \eta)}},\label{eq:msecondybtbound}
\eeqa
where $\zeta > 0$ is an arbitrary number.
In the above, \emph{(a)} and \emph{(b)} follow from the Markov inequality and Jensen's inequality, respectively, and \emph{(c)} follows from
Lemma \ref{lem:mmseprop}. Since $\eta < 1$, and $\zeta$ can be chosen arbitrarily large, the second term in \eqref{eq:poutexpsecondterm} goes to zero with an arbitrarily large exponent as $\barp$ goes to infinity.

Putting \eqref{eq:goodbad1}, \eqref{eq:good2}, \eqref{eq:bad2} and \eqref{eq:msecondybtbound} together, a DMT of $d(g_m) = r\left(\min\{l,s\}  + \eta- \frac{g_m}{\alpha}\right)$
is achievable, for $0 \leq l \leq r$, $1 \le s < r$ and $0 \le g_m < \eta \alpha$. Noting that $\eta$ is arbitrarily close to $1$ completes the proof of Theorem \ref{thm:dmttwoway}. $\blacksquare$

\section*{Acknowledgments}
The authors thank the anonymous reviewers for their detailed comments and suggestions, which have improved the quality of this paper. 

\end{document}